%% file: arxiv/main.tex
\documentclass{article}
\usepackage[utf8]{inputenc}
\usepackage[english]{babel}

\usepackage[letterpaper,top=2cm,bottom=2cm,left=3cm,right=3cm,marginparwidth=1.75cm]{geometry}

\usepackage{amsmath,amssymb}
\usepackage{braket}
\usepackage{graphicx}
\usepackage{algorithm}
\usepackage[colorlinks=true, allcolors=blue]{hyperref}
\usepackage{authblk}
\usepackage[toc,page]{appendix}
\usepackage{listings}
\usepackage{xcolor}
\usepackage{tabularx}
\usepackage{multicol}
\usepackage[htt]{hyphenat}
\usepackage[hyphenbreaks]{breakurl}
\usepackage{textcomp}
\usepackage{stmaryrd}
\usepackage{diagbox}
\usepackage{verbatim}

\definecolor{codecomment}{RGB}{64, 128, 128}
\definecolor{codegreen}{RGB}{0, 128, 0}
\definecolor{backcolour}{RGB}{247, 247, 247}
\definecolor{codered}{RGB}{186, 33, 33}
\lstdefinestyle{python}{
    backgroundcolor=\color{backcolour},   
    commentstyle=\color{codecomment},
    keywordstyle=\color{codegreen},
    stringstyle=\color{codered},
    basicstyle=\ttfamily\footnotesize,
}

\DeclareUnicodeCharacter{211D}{$\mathbb{R}$}

\title{Factorizing binary tensors into quantics tensor trains}
\author[a,b]{Paul Haubenwallner\footnote{\href{mailto:paul.haubenwallner@igd.fraunhofer.de}{paul.haubenwallner@igd.fraunhofer.de}}}
\author[a,b]{Matthias Heller\footnote{\href{mailto:matthias.heller@igd.fraunhofer.de}{matthias.heller@igd.fraunhofer.de}}}
\affil[a]{Fraunhofer Institute for Computer Graphics Research IGD, Darmstadt, Germany}
\affil[b]{Technical University of Darmstadt, Interactive Graphics Systems Group, Darmstadt, Germany
}
\date{}

\begin{document}

\maketitle
\vspace{-1cm}

\begin{abstract}
\input{sections/abstract}
\end{abstract}

% \tableofcontents

\input{sections/introduction}
\input{sections/notation}
\input{sections/general_algorithm}
\input{sections/applications}
\input{sections/conclusion}

\section*{Acknowledgements}
This work was supported by the research project ``Zentrum für Angewandtes Quantencomputing'' (ZAQC), which is funded by the Hessian Ministry for Digital Strategy and Innovation and the Hessian Ministry of Higher Education, Research and the Arts.
We especially thank Timon Scheiber for interesting discussions and constructive feedback.

\clearpage

\bibliographystyle{ieeetr}
\bibliography{references}

\end{document}

%% file: sections/abstract.tex
The conversion of functions to quantics tensor trains is a well-established procedure and can either be done analytically or numerically.
Numerical conversion schemes are based on singular value decompositions, where access to the full tensor is necessary, or on cross interpolations, which only depend on sampling a function.
When dealing with large binary tensors, the first approach becomes prohibitively expensive while the second approach might fail to converge due to the non-smoothness of the data.
In this work, we provide insight into how binary tensors, where the positions of the non-zero entries are defined by some Boolean function, can be converted into quantics tensor trains with a hybrid analytical-numerical approach utilizing the rank product.
The proposed construction scheme nicely reproduces results from the literature and can be used for discrete convolutions, the construction of discrete wavelet transforms or slicing and assignment operations of multi-dimensional quantics tensor  trains.

%% file: sections/introduction.tex
\section{Introduction} \label{sec:introduction}

Binary tensors are multi-dimensional arrays where the entries are either zero or one.
They are often encountered in scientific computing especially when dealing with Boolean or indexing operations~\cite{ArrayStandard}.
Although rarely done, they can be used as a basic construct for other linear algebra operations as well, like numerical differentiation~\cite{Kazeev_2012} or discrete convolutions~\cite{Kazeev2013}.
The reason for their infrequent use is that the effect of many of those binary tensors can be taken into account implicitly.
A good example is the calculation of the second derivative using finite differences~\cite{Danaila_2023}.
The idiomatic way is to write down the full matrix, which is a sum of three shift matrices multiplied by some coefficient, and perform a matrix multiplication with a vector to calculate its second derivative.
It is, however, much more efficient to perform the operation implicitly by multiplying and adding only adjacent neighbors avoiding the operations on values known to be zero.

When dealing with compressed data in the form of some kind of tensor network~\cite{Banuls_2023, Ritter_2024, Berezutskii_2025}, it is generally not possible to implicitly take the effects of binary tensors into account, so that the tensor has to be explicitly applied.
The construction of such a binary tree-like tensor network, in this work into a tensor train, can either be done analytically or numerically.
For some tensors the analytical factorization is known, with shift and Toeplitz matrices as the most prominent examples~\cite{Kazeev2013}.
An important advantage of an analytical factorization is the speed with which it can be carried out.
Once derived, it is only necessary to allocate and fill the tensors of the tensor network according to some formula.
In contrast to that, numerical approaches tend to scale poorly or might not converge at all.
The first point is attributed to factorizations based on singular value decompositions~\cite{Khoromskij_2011}, where the full tensor is constructed and decomposed.
This approach is of course infeasible for exponentially large tensors but tends to provide good approximations.
The second point is attributed to cross interpolations~\cite{Oseledets_2010,Savostyanov_2011,Savostyanov_2014}, which work out of the box for smooth data, but have problems with functions containing discontinuities.
Since important binary tensors can be very sparse and cross approximations are based on sampling a function, they often miss out on important features of the tensor and therefore cannot efficiently reproduce the structure.
This problem can, however, be counteracted by choosing global pivots as described in~\cite{NunezFernandez_2025}.

In this work, we provide general rules on how binary tensors, in which the positions of the non-zero entries are described by some Boolean function, can be exactly factorized and how the rank can be kept low during this factorization.
We start with an introduction to our notation in Section~\ref{sec:notation}, followed by the proposed factorization scheme in Section ~\ref{sec:general_algorithm}.
Section~\ref{sec:applications} contains multiple applications making use of the derived scheme together with numerical examples.
Finally, in Section~\ref{sec:conclusion}, we conclude the work.

%% file: sections/notation.tex
\section{Notation} \label{sec:notation}

\newcommand{\liv}[0]{\llbracket}
\newcommand{\riv}[0]{\rrbracket}
\newcommand{\xdim}[1]{x^{(#1)}}
\newcommand{\coeff}[2]{#1^{(#2)}}
\newcommand{\xpara}[0]{\xdim{1},\dots,\xdim{N}}

%\paragraph{Tensor trains}
Tensor networks are sets of multiple tensors that are contracted with each other by some predefined order.
We denote dimensions that are contracted by the network as $\mu$ (often referred to as bond dimension) and those that are not as a lower case letter with an index $x_i$.
Tensors are denoted as capital letters with the dimensions in parentheses, such that $A(x_i)$ is a tensor with one dimension $x_i$, where $x_i \in \{0,1,\dots, \text{dim}(x_i)-1\}$.
Using these definitions, tensor trains are a way of representing $N$-dimensional tensors $A(x_1,x_2,\dots,x_N)$ as linear tensor networks:
\begin{equation}
    A(x_1,\dots,x_N) \approx \sum_{\mu_1\dots,\mu_{n+1}} A_1(\mu_1, x_1,\mu_2) A_2(\mu_2,x_2,\mu_3)\dots A_n(\mu_n,x_N,\mu_{n+1}).
    \label{eq:tt_decomposition}
\end{equation}
The number of cores is arbitrary and any assignment of the dimensions $x_n$ to a core $A_n$ is valid as long as each core has at least one dimension $x_n$ associated with it.
For example, the decomposition
\begin{equation}
    A(x_1,\dots,x_N) \approx \sum_{\mu_1\dots,\mu_{n}} A_1(\mu_1, x_2, x_3,\mu_2) A_2(\mu_2,x_1,\mu_3) A_3(\mu_3,x_4,\mu_4)\dots A_n(\mu_{n-1},x_N,\mu_{n}).
    \label{eq:arb_order}
\end{equation}
represents a valid linear tensor network.
Another way of expressing tensor trains is by utilizing the so-called rank product $\boxtimes$, which is defined by
\begin{align}
    &    C(\mu_1,x_i,\mu_2)\boxtimes D(\mu_2,x_j,\mu_3) \\[1.5em]
    &=\left(\begin{array}{ccc}
        C(0,x_i,0) & \dotsc & C(0,x_i,M_2) \\
        \vdots &  & \vdots \\
        C(M_1,x_i,0) & \dotsc & C(M_1,x_i,M_2) \\
    \end{array}\right)\boxtimes
    \left(\begin{array}{ccc}
        D(0,x_j,0) & \dotsc & D(0,x_j,M_3) \\
        \vdots &  & \vdots \\
        D(M_2,x_j,0) & \dotsc & D(M_2,x_j,M_3) \\
    \end{array}\right)\nonumber\\[1.5em]
    &=\left(\begin{array}{ccc}
        \sum^{M_2}_{m=0} C(0,x_i,m)\otimes D(m,x_j,0) & \dotsc & \sum^{M_2}_{m=0} C(0,x_i,m)\otimes D(m,x_j,M_3) \\
        \vdots &  & \vdots \\
        \sum^{M_2}_{m=0} C(M_1,x_i,m)\otimes D(m,x_j,0) & \dotsc & \sum^{M_2}_{m=0} C(M_1,x_i,m)\otimes D(m,x_j,M_3) \\
    \end{array}\right),
\end{align}
where $M_n=\text{dim}(\mu_n)-1$ and $\otimes$ denotes the usual outer product of tensors.
It can be used to rewrite expression~\ref{eq:tt_decomposition} as
\begin{align}
    &\left(\begin{array}{c}
        A_1(0,x_1,0) \\
        A_1(0,x_1,1)\\
        \vdots \\
        A_1(0,x_1,M_2)\\
    \end{array}\right)^T \boxtimes  \left(\begin{array}{ccc}
        A_2(0,x_2,0) & \dotsc & A_2(0,x_2,M_3) \\
        \vdots & & \vdots \\
        A_2(M_2,x_2,0) & \dotsc & A_2(M_2,x_2,M_3) \\
\end{array}\right)\boxtimes\cdots\boxtimes \left(\begin{array}{c}
        A_n(0,x_n,0) \\
        A_n(1,x_n,0)\\
        \vdots \\
        A_n(M_n,x_n,0)\\
    \end{array}\right).
    \label{eq:rank_prod}
\end{align}
The quantics representation of a dimension $x_i$ is defined by the factorization
\begin{equation}
x_i=\sum^n_{q=1} \left(\prod_{r={q+1}}^n b_{i,r}\right)\cdot x_{i,q}\equiv \sum^n_{q=1} c_{i,q} x_{i,q},
\label{eq:quantization}
\end{equation}
where $c_{i,q}=\prod_{r={q+1}}^n b_{i,r}$ and the integer bases $b_{i,r}$ are chosen such that
\begin{equation}    
{\rm dim}(x_i) = c_{i,0} = \prod^n_{q=1} b_{i,q}.
\end{equation}
With the decomposition of $x_i$, $A(x_i)$ becomes $A(x_{i,1},\dots,x_{i, n_i})$, which corresponds to reshaping the tensor into another one with more, but lower-dimensional indices.
In the following, we will refer to this process as factorizing a dimension (or index).
Quantics tensor trains are defined by representing the full tensor $A(x_{i,1},\dots,x_{i, n})$ according to equation~\ref{eq:tt_decomposition}.
This kind of decomposition is of course also possible for multiple dimensions, so that for example a matrix $A(x_1,x_2)$ is transformed into $A(x_{1,1},x_{1,2},\dots,x_{2,1},x_{2,2},\dots)$ and then approximated as tensor train.

A hyperplane of an $N$-dimensional tensor (or function) $A(x_1,\dots,x_N)$ is a slice of this tensor along one or multiple dimensions.
Let $I \subset \{1,\dots,N\}$ be the set of indices of the dimensions that are fixed.
We write $\tilde{x}=(x_i)_{i\notin I}$ for the indexed family of all unfixed indices, and $\dot{x}=(\dot{x})_{i\in I}$ for the family, where each $\dot{x}_i$ is a chosen value of $x_i$ for $i\in I$.
For later reference we also define $\hat{x}_{i\in I}$, which is the family of the indices in $I$ viewed as variables.
The slice $B$, obtained by fixing $x_i=\dot{x}_i$ for all $i\in I$, is then the function of the remaining variables given by
\begin{equation}
    B(\tilde{x}) = A(\tilde{x}, \dot{x}).
    \label{eq:slice}
\end{equation}
As an example, consider $A(x_1,x_2,x_3)$ with $\dim(x_1)=\dim(x_2)=\dim(x_3)=10$, and fix $x_1=5$.
In this case $I=\{1\}$, so $\tilde{x}=(x_i)_{i\in \{2,3\}}$, $\hat{x}=(\hat{x})_{i\in \{1\}}$ and $\dot{x}=(\dot{x}_1)_{i\in \{1\}}$ with $\dot{x}_1=5$. The resulting slice is
\begin{equation}
    B(x_2, x_3) = A(x_1=5,x_2,x_3),
\end{equation}
or with the introduced notation equivalently $B(\tilde{x}) = A(\tilde{x}, \dot{x})$.

A shorthand notation for binary tensors $T(x_1,\dots,x_N)$ is the Iverson bracket, which we will depict as $\liv...\riv$.
It is defined by
\begin{equation}
    T(x_1,\dots,x_N) = \left\liv f(x_1,...,x_N) \right\riv = \begin{cases}
        1 \mbox{ if } f(x_1,...,x_N)=\text{true},\\
        0 \mbox{ otherwise}.
    \end{cases}
\end{equation}
where $f$ is some Boolean function that takes an index tuple and evaluates it to true or false.

%% file: sections/general_algorithm.tex
\newcommand{\mtup}[0]{\mathbf{m}}
\newcommand{\ntup}[0]{\mathbf{d}}
\newcommand{\Ntup}[0]{D}

\newcommand{\fset}[0]{I}
\newcommand{\xo}[0]{\tilde{x}}
\newcommand{\xt}[0]{\hat{x}}

\section{Factorization scheme}\label{sec:general_algorithm}

\begin{figure}
    \centering
    \includegraphics[width=\linewidth]{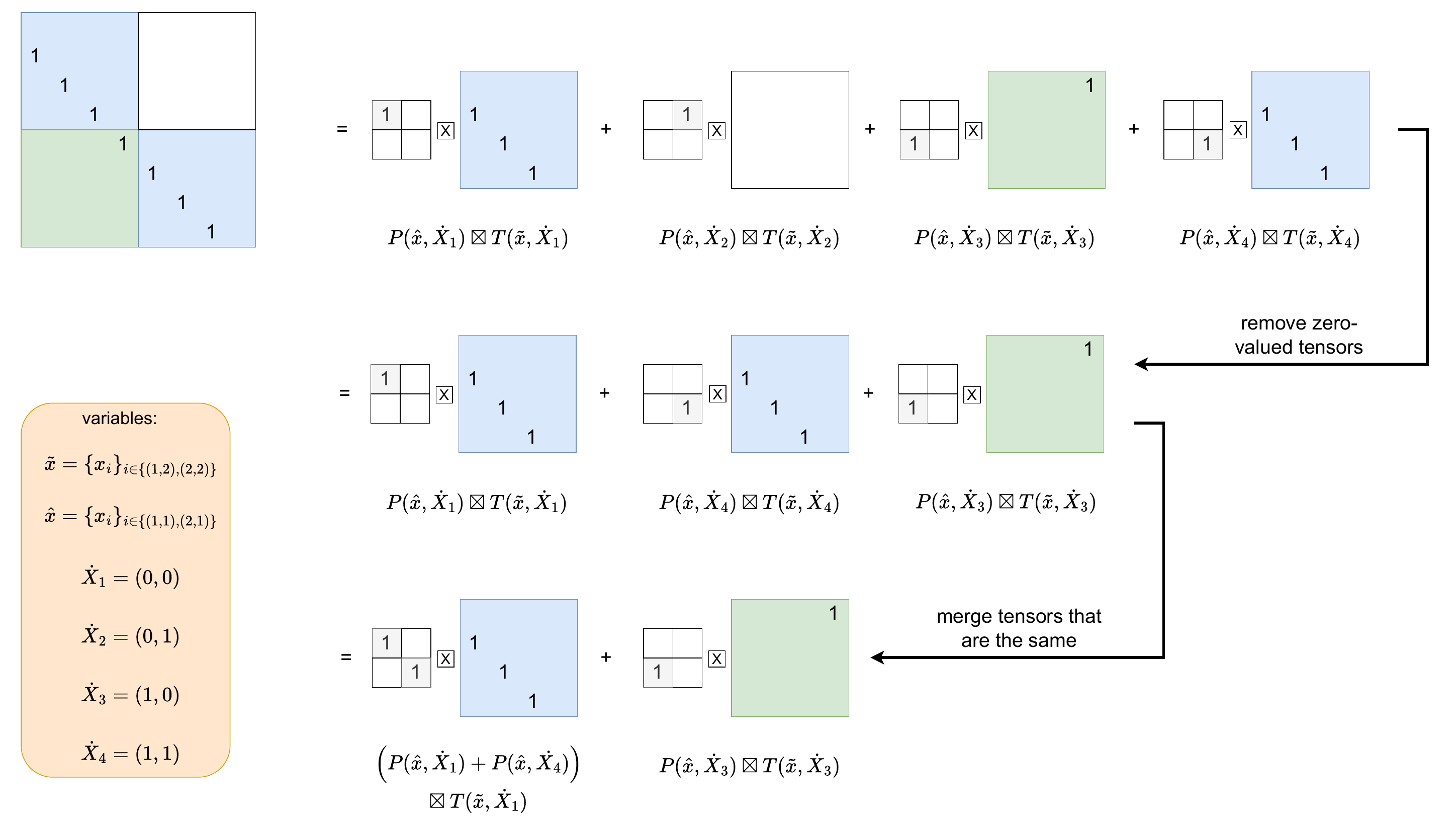}
    \caption{Graphical example for a single factorization step in quantics representation of a $(8, 8)$ shift matrix defined by $F(x_{1,1},x_{1,2},x_{2,1},x_{2,2})=\left\liv 4x_{1,1}+x_{1,2}-1=4x_{2,1}+x_{2,2} \right\riv$ according to \ref{eq:general_factorization} with subsequent rank reduction steps.}
    \label{fig:shift_example}
\end{figure}

The goal of this section is to show how to factorize an $N$-dimensional binary tensor $T(x_1,\dots,x_N)$, defined by a Boolean function $f$
\begin{equation}
    T(x_1,\dots,x_N) = \left\liv f(x_1,\dots,x_N) \right\riv,
    \label{eq:binary_tensor}
\end{equation}
into a tensor train.
%The construction is based on the fact, that any tensor can be easily represented by utilizing its own hyperplanes.
Let $I$ be the set, which holds the indices of those dimensions $x_i$, that we want to factor out of the binary tensor $T$.
Let $\dot{X}$ be the tuple of all possible assignments to the index set $(x_i)_{i\in I}$ ordered lexicographically with size $Q=\prod_{i\in I}\text{dim}(x_i)$.
Using the notation for hyperplanes introduced in Section~\ref{sec:notation}, any binary tensor $T(x_1,\dots,x_N)$ can be decomposed into
\begin{equation}
    T(x_1,\dots,x_N) = \sum^Q_{q=1} P(\hat{x}, \dot{X}_q) \otimes T(\tilde{x},\dot{X}_q),
    \label{eq:sum_factorization}
\end{equation}
where
\begin{equation}
    P(\hat{x},\dot{X}_q) = \left\liv\hat{x}=\dot{X}_q \right\riv.
\end{equation}
Here, each hyperplane $T(\tilde{x}, \dot{X}_q)$ is a slice of the original tensor at the fixed position $\hat{x}=\dot{X}_q$.
The tensor $P(\hat{x}, \dot{X}_q)$ acts as a ``selector'' that places this hyperplane at the correct position in the full tensor and sets all other positions to zero.
Summing over all $q$ reconstructs $T$ exactly.
To connect this type of factorization with the concept of a tensor train, we can rewrite equation \ref{eq:sum_factorization} in terms of the rank product:
\begin{equation}
    T(x_1,\dots,x_N) = \left(\begin{array}{cccc}
        P(\hat{x},\dot{X}_1) & P(\hat{x},\dot{X}_2) & \cdots & P(\hat{x},\dot{X}_Q)
    \end{array}\right) \boxtimes \left(\begin{array}{c}
        T(\tilde{x},\dot{X}_1) \\
        T(\tilde{x},\dot{X}_2) \\
        \vdots \\
        T(\tilde{x},\dot{X}_Q) \\
    \end{array}\right).
    \label{eq:single_factorization}
\end{equation}
The row vector collects all selector tensors $P(\hat{x}, \dot{X}_q)$; the column vector collects the corresponding hyperplanes $T(\tilde{x}, \dot{X}_q)$.
The rank introduced at this factorization step is thus $Q$.
We can abbreviate equation~\ref{eq:single_factorization} as
\begin{equation}
    T(x_1,\dots,x_N) = \mathbf{P}(\hat{x}) \boxtimes \mathbf{T}(\tilde{x}),
\end{equation}
with $\mathbf{P}(\hat{x})$ as the row vector of all $P(\hat{x},\dot{X}_q)$ and $\mathbf{T}(\tilde{x})$ as the column vector of all $T(\tilde{x}, \dot{X}_q)$.
A visual depiction of this kind of decomposition can be seen in the first line of Figure~\ref{fig:shift_example}.

For further factorization steps we must be able to apply the same decomposition not just to a single binary tensor, but to a vector of binary tensors of the same shape. This is straightforward: we apply the above decomposition to each entry separately.
Let $(T_r (x_1,\dots,x_N))_{(r=1)}^R$ be such a vector.
Then we can write
\begin{equation}
    \left( \begin{array}{c}
         T_1(x_1,\dots,x_N)  \\
         \vdots  \\
         T_R(x_1,\dots,x_N)  
    \end{array} \right) = \left( \begin{array}{ccc}
        \mathbf{P}_1(\hat{x}) &  & \\
         & \ddots & \\
         &  & \mathbf{P}_R(\hat{x})
    \end{array} \right) \boxtimes \left( \begin{array}{c}
         \mathbf{T}_1(\tilde{x})  \\
         \vdots  \\
         \mathbf{T}_R(\tilde{x})  \\
    \end{array} \right).
    \label{eq:general_factorization}
\end{equation}
Here, the $P$-matrix is block diagonal so that each block is the row vector $\mathbf{P}_r(\hat{x})$.
The vector on the right is simply the stack of the corresponding $\mathbf{T}_r(\tilde{x})$ columns.
As before, the rank of this factorization step equals the length of the right-hand vector.
Applying equation~\ref{eq:general_factorization} recursively starting with a single $T$ allows for the exact factorization of arbitrary binary tensors.
The $P$-matrices together with the $T$-vector of the last step can be used to define the cores of a tensor train.
As the rank of each factorization depends on the rank of the last factorization, the maximum rank of such a tensor train increases exponentially with the number and size of the dimensions of the binary tensors.
For highly structured tensors, however, this growth of the rank can be counteracted at each factorization step by exploiting structural redundancy.

One strategy for reducing the rank is based on identifying those hyperplanes, which have only zero as entries.
For each $q$, consider the Boolean condition generating $T(\tilde{x},\dot{X}_q)$, i.e., the equation $f(\tilde{x},\dot{X}_q)=\text{true}$ .
If this equation has no solution in the finite domain of $\tilde{x}$, then the tensor $T(\tilde{x},\dot{X}_q)$ is zero along the entire hyperplane and can be removed from the decomposition, which reduces the rank by one for each such case.
Since the domain is finite, this check for ``solvability'' can always be done by brute force evaluation of $f$ if necessary.
However, for certain function classes, the check can be done much more efficiently than in the general case.

The second strategy relies on detecting linear dependencies between different hyperplanes, i.e., one has to check if there exist non-trivial $\lambda_i$ such that
\begin{equation}
    \sum_q \lambda_q T(\tilde{x},\dot{X}_q) = 0.
\end{equation}
The simplest case occurs when some hyperplanes are identical.
More generally, we consider the Gram-matrix
\begin{equation}
    G(q,r)=\braket{T(\tilde{x},\dot{X}_q)|T(\tilde{x},\dot{X}_r)},
\end{equation}
where $T(\tilde{x},\dot{X}_q)$ are treated as the basis vectors.
The rank of $G$ equals the number of linearly independent hyperplanes.
We select an index subset $S$ such that $\{T(\tilde{x},\dot{X}_q)\}_{q\in S}$ is linearly independent; equivalently, the principal submatrix $G(q\in S, r\in S)$ has full rank.
Every hyperplane can then be expressed as a linear combination of this basis.
If 
\begin{equation}
    \sum_{j\in S}C(i,j\in S) G(j\in S,k\in S) = G(i,k\in S),
\end{equation}
the rows of the transformation matrix $C$ encode the coefficients expressing each $T(\tilde{x},\dot{X}_i)$ in the basis $\{T(\tilde{x},\dot{X}_q)\}_{q\in S}$.
Multiplying the original $P$-matrix from the left by $C$ yields a new $P$-matrix of reduced rank that is still sufficient to represent equation~\ref{eq:general_factorization} exactly, while the actual set of basis hyperplanes remains unchanged.

A visualization for a single factorization step with subsequent rank reduction is depicted in Figure~\ref{fig:shift_example}.
Before continuing, it is very important to stress that the described factorization does not rely on a successful rank reduction.
Even if empty hyperplanes are kept or some linear dependencies are not found, the correctness of the factorization itself is unaffected.
In the following paragraphs we will investigate three groups of Boolean functions that allow efficient rank reduction strategies.

\paragraph{Linear functions} \label{sec:lin_funcs}
We consider $N$-dimensional binary tensors of the form
\begin{equation}
    G(x_1,\dots,x_N) = \left\liv \sum^N_{i=1} c_ix_i = \Delta \right\riv,
    \label{eq:linear_equation}
\end{equation}
with integer coefficients $c_i$ and the right-hand side $\Delta$.
Following the notation of section~\ref{sec:notation} we can express any hyperplane defined by the set of the fixed dimensions $I$ with
\begin{equation}
    G(\tilde{x}, \dot{x}) = \left\liv \sum_{i\notin I} c_i x_i = \Delta -  \sum_{i\in I} c_i \dot{x}_i \right\riv.
\end{equation}
Introducing the effective right-hand side
\begin{equation}
    \tilde{\Delta} = \Delta -  \sum_{i\in I} c_i \dot{x}_i,
\end{equation}
which is constant for a given index assignment, the decision whether a hyperplane of the corresponding binary tensor contains any non-zero entries reduces to checking feasibility of the integer linear program (ILP)
\begin{equation}
    \sum_{i\notin I} c_i x_i - \tilde{\Delta} = 0,\qquad
    0 \leq x_i < \dim(x_i).
\end{equation}
If a feasible solution exists, the hyperplane contains at least one non-zero entry, otherwise it is identically zero.
Linear dependencies between hyperplanes arising from different index assignments in $\dot{X}$ can be detected directly from $\tilde{\Delta}$: All hyperplanes with the same effective right-hand side share the same solution set and thus correspond to identical binary tensors.
Hyperplanes with different values of $\tilde{\Delta}$ have disjoint solution sets on the finite index domain, and their indicator tensors are linearly independent.

Using the quantics formulation we can rewrite equation~\ref{eq:linear_equation} as
\begin{equation}
    G(x_1,\dots,x_N) = \left\liv \sum^N_{i=1} \sum^{n_i}_{q=1} c_{i,q} x_{i,q} = \Delta \right\riv.
\end{equation}
Hence, in quantics form the defining relation remains linear in the new indices $x_{i,q}$.
All hyperplanes of the quantics tensor are again described by linear constraints of the same type, so the feasibility and redundancy checks above apply unchanged.
Consequently, binary tensors defined by linear functions can be systematically converted into quantics tensor trains using this factorization scheme.

\paragraph{Modulo functions} \label{sec:mod_funcs}
We next consider binary tensors of the form
\begin{equation}
    H(x_1,\dots,x_N) = \left\liv \sum_{i=1}^N (x_i + o_i) \mod \delta_i = \Delta \right\riv,
    \label{eq:modulo}
\end{equation}
with the moduli $\delta_i\in \mathbb{N}$, offsets $o_i\in \mathbb{Z}$ and a target value $\Delta\in \{0, 1, \dots, \sum_{i=1}^N(\delta_i-1)\}$.
An arbitrary hyperplane with the fixed dimensions $I$ is defined by
\begin{equation}
    H(\tilde{x}, \dot{x}) = \left\liv \left(\sum_{i\notin I} (x_i + o_i) \mod \delta_i\right) + \left(\sum_{i\in I} (\dot{x}_i + o_i) \mod \delta_i\right) = \Delta \right\riv
\end{equation}
To decide whether this hyperplane contains any non-zero entries, we formulate a small integer-linear feasibility problem.
For every unfixed dimension $i\notin I$  we introduce quotient and remainder variables $k_i\in Z$ and $r_i\in \{0,\dots,\delta_i-1\}$ and define the ILP:
\begin{equation}
    x_i + o_i = \delta_i k_i + r_i,
    \label{eq:modulo_ilp1}
\end{equation}
with the bound on $k_i$ induced by $0\leq x_i<\dim(x_i)$.
Introducing the effective right hand side $\tilde{\Delta}$
\begin{equation}
    \tilde{\Delta} = \Delta - \sum_{i\in I} (\dot{x}_i + o_i) \mod \delta_i,
\end{equation}
the modular constraint becomes
\begin{equation}
    \sum_{i\notin I} r_i = \tilde{\Delta}.
    \label{eq:modulo_ilp2}
\end{equation}
If the system defined by equations \eqref{eq:modulo_ilp1}-\eqref{eq:modulo_ilp2} has no solution, the corresponding hyperplane is identically zero.
Linear dependencies between hyperplanes can again be detected from $\tilde{\Delta}$ and $o_i$: all hyperplanes with the same values for $\tilde{\Delta}$ and $o_i\mod{\delta_i}$ coincide, while hyperplanes with different values are linearly independent on the finite index domain.

Inserting equation~\eqref{eq:quantization} into \eqref{eq:modulo} yields
\begin{equation}
    H(x_1,\dots,x_N) = \left\liv \sum_{i=1}^N \left(\sum^{n_i}_{q=1} c_{i,q} x_{i,q} + o_i\right) \mod \delta_i = \Delta \right\riv.
\end{equation}
Thus, in quantics form we obtain modular equations of the same type, now in the lower dimensional indices $x_{i,q}$.
The ILP-based feasibility test and the rank-reduction strategy carry over directly.
The only refinement is that contributions from fixed quantics indices need to be absorbed into the offsets, so that linear dependencies can again be identified by comparing the resulting effective right-hand sides and modified $o_i$.

\paragraph{Range-based functions} \label{sec:range_funcs}
A third class of binary tensors is given by Boolean functions with axis-aligned range constraints on each index.
We consider
\begin{equation}
    F(x_1,\dots,x_N) = \prod_{i=1}^N \left\liv l_i \leq x_i < u_i \right\riv,
\end{equation}
where $l_i,\ u_i\in \mathbb{Z}$ with $0\leq l_i<u_i\le\dim(x_i)$.
A hyperplane of this tensor is defined by the fixed dimensions $I$ and their corresponding offsets $\dot{x}$
\begin{equation}
    F(\tilde{x},\dot{x}) = \prod_{i\notin I} \left\liv l_i \leq x_i < u_i \right\riv \prod_{i\in I} \left\liv l_i \leq \dot{x}_i < u_i \right\riv.
\end{equation}
To decide whether this hyperplane is non-zero, we first check if all fixed indices satisfy their range constraints:
\begin{equation}
    l_i \leq \dot{x}_i < u_i \ \ \ \text{for all } i\in I.
\end{equation}
If this condition fails for at least one $i\in I$ , the entire hyperplane is identically zero.
For each unfixed dimension $i\notin I$ we need to test if the interval $[l_i,u_i)$ intersects with $x_i$.
This can be formulated as a pair of small integer linear programs per dimension: one minimizes and one maximizes $x_i\notin I$ under the constraint
\begin{equation}
    l_i \leq x_i < u_i.
\end{equation}
If either the optimization problem is infeasible or the intersection is empty, the corresponding hyperplane is identically zero.
The inner product of two hyperplanes can be calculated as the product of the distances between the minimized and maximized values.
Hence, the Gram matrix can be computed explicitly, which allows one to detect linear dependencies as described earlier.

Using the quantics representation of the problem
\begin{equation}
    F(x_1,\dots,x_N) = \prod_{i=1}^N \left\liv l_i \leq \sum_{q=1}^{n_i} c_{i,q} x_{i,q} < u_i \right\riv,
\end{equation}
the overall solutions stay the same.
When calculating the inner product of different hyperplanes, however, it is necessary to consider that the index blocks might be not contiguous anymore.
Consequently, when computing inner products between hyperplanes (and thus the Gram matrix), one must count the entries of these disjoint blocks rather than relying on a single contiguous interval length.

%% file: sections/applications.tex
\section{Applications} \label{sec:applications}
In the following we will show how the aforementioned functions can be applied for several important problems when dealing with quantics tensor trains.
Unless otherwise indicated, numerical examples were carried out by providing a predefined order of the integer bases $x_{i,q}$ and factors $c_{i,q}$, but, in the multi-dimensional cases, without defining the orders of each dimension with respect to each other.
In these cases we used a local greedy search, which minimizes the rank at every step with the prerequisite, that at least one dimension is factorized.
The numerical examples have been implemented in and executed with the Python package \texttt{trainsum}~\cite{Haubenwallner2026}.

\paragraph{Reordering indices of quantics tensor trains}
A quantics tensor train has, in analogy to a normal tensor, a certain shape.
The shape is, however, more complex since the dimensions of the approximated tensor have been factorized, so that it can be described as a nested sequence of the factorized indices (equation \ref{eq:quantization}).
The order of the factorized indices can be completely arbitrary, yet some orders lead to a low-rank approximation while others do not.
A perfect example for this kind of behavior is the discrete Fourier transformation, where a tensorization is only possible if the factorized indices of rows and columns are arranged in opposite directions \cite{Chen_2023}.
\begin{figure}
    \centering
    \includegraphics[width=0.92\linewidth]{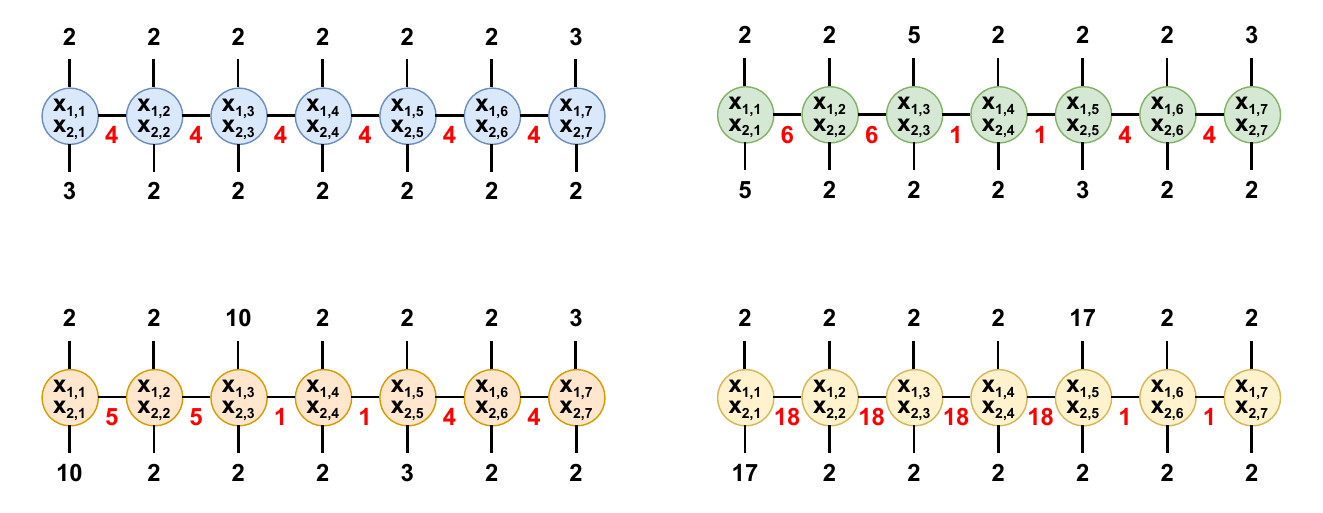}
    \caption{Graphical depiction (Penrose) of the tensor trains, that have been factorized according to equation \ref{eq:general_factorization} for the function $f(x_1,x_2)=\left\liv x_1=x_2 \right\riv$. The varying ranks indicate that the internal structure has been unveiled for the displayed shape of the tensor train. Numerical compression schemes lead to the same ranks, therefore supporting the findings.}
    \label{fig:reorder}
\end{figure}

If some data is approximated by a quantics tensor train it is easily possible to change the meaning of the cores by applying SWAP operations, which also often arise in the context of quantum computing.
A SWAP operation swaps two factorized indices and is known to have a rank of 4.
Another similar operation can be formulated with the function
\begin{equation}
    f(x_1,x_2) = \left\liv x_1=x_2 \right\riv,
\end{equation}
which essentially defines an identity matrix.
With $\text{dim}(x_1)=\text{dim}(x_2)$ and different $c_{1,q}$ and $c_{2,q}$ during the construction process of the tensor train, it is possible to map the dimension $x_2$ to the dimension $x_1$, which has a different factorization, while still describing the same data.
Numerical examples for this kind of operators are graphically depicted in figure \ref{fig:reorder}.

\paragraph{Multi-dimensional discrete convolutions}
The application of binary tensors to describe multi-dimensional discrete convolutions has been already shown in Ref.~\cite{Kazeev2013}.
We merely want to note here that the corresponding binary tensors for one-dimensional convolutions more generally take forms like
\begin{equation}
    C(x_1,x_2,x_3) = \left\liv c_{x_1} x_1 + c_{x_2} x_2 + c_{x_3} x_3 = \Delta \right\riv,
\end{equation}
which allows to convolve vectors with different sizes and arbitrary decomposition of their dimensions.

\paragraph{Slicing and assignment operations of quantics tensor trains}
\newcommand{\slstart}[0]{\mu}
\newcommand{\slstop}[0]{\nu}
\newcommand{\slstep}[0]{\gamma}
Slicing operations are indexing operations that select specific elements of a tensor and create a new tensor with this selection, while retaining the number of (then smaller) dimensions.
The selection indices of a single dimension can be expressed as so-called slices, which are defined by a start value $\slstart$, a stop value $\slstop$ and a step value $\slstep$, so that they describe
\begin{equation}
    (i_1,i_2,...,i_M) = (\slstart+m\times\slstep)_{m=0}^{M-1},
    \label{eq:slice}
\end{equation}
where $M=\left\lfloor \frac{\slstop-\slstart}{\slstep} \right\rfloor$.
\begin{figure}[h!]
    \centering
    \includegraphics[width=0.9\linewidth]{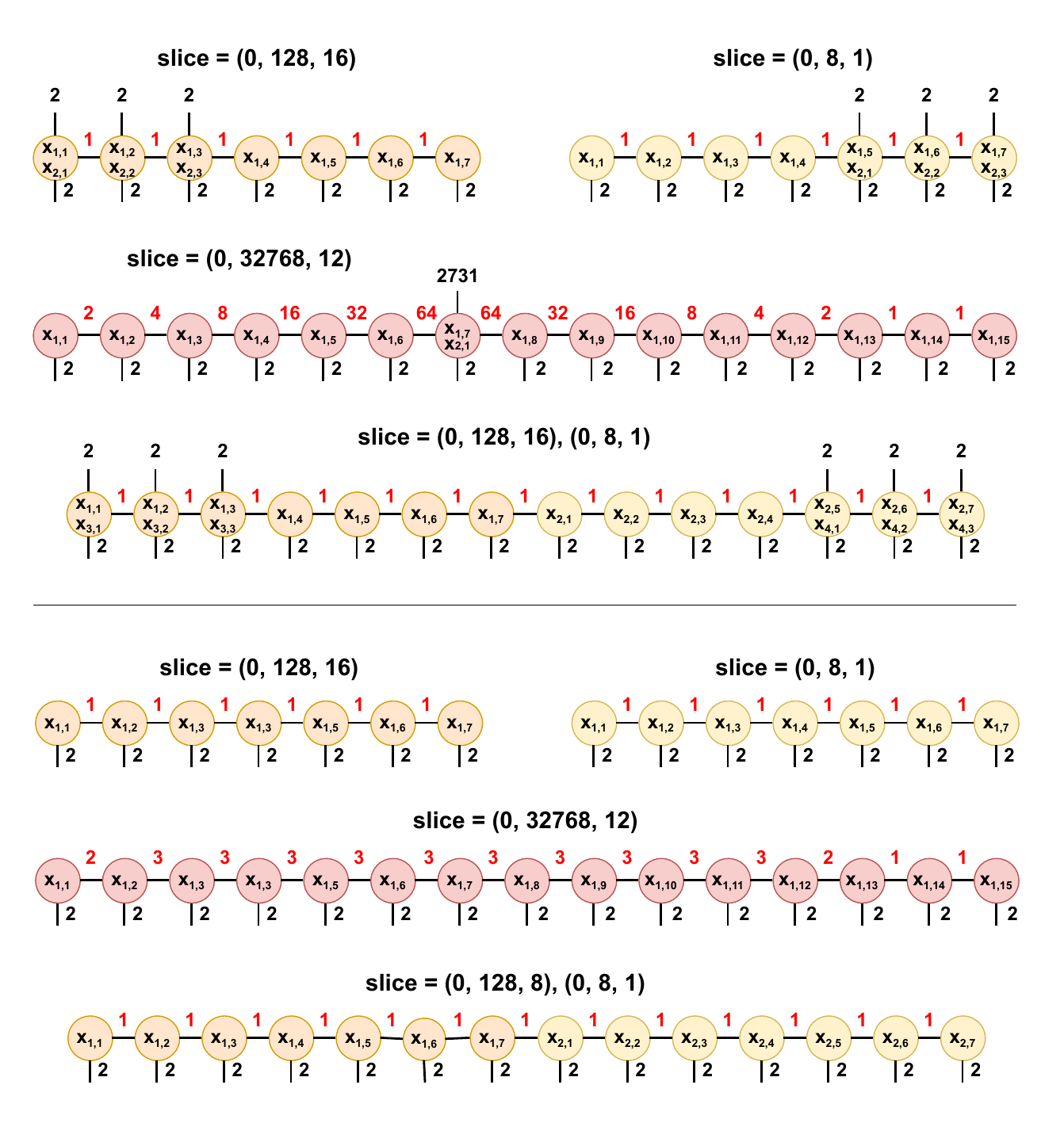}
    \caption{Graphical depiction of the tensor trains, which represent equation \ref{eq:slice_eq} (top) and equation \ref{eq:slice_inverse} (bottom), where a slice is defined by $(\slstart, \slstop, \slstep)$. As can be seen, the rank heavily depends on which slice is represented. The most efficient mapping are found for slices that match the factorization of the tensor train. Multi-dimensional slices can be represented as the outer product. All findings are in accordance with numerical experiments using an SVD-decomposition scheme for converting the full matrices and tensors to tensor trains.}
    \label{fig:slicing}
\end{figure}
Using the binary tensor described by the linear equation
\begin{equation}
    S(x_1,x_2) = \left\liv \slstep\ x_1 = x_2-\slstart \right\riv,
    \label{eq:slice_eq}
\end{equation}
with $\text{dim}(x_1)=\left\lfloor \frac{\slstop-\slstart}{\slstep} \right\rfloor$, it is possible to describe a one-dimensional slicing operation as the summation
\begin{equation}
    B(x_1) = \sum_{x_2} S(x_1,x_2) A(x_2),
\end{equation}
where the values of $A(x_2)$ defined by the slice $(\slstart,\slstop,\slstep)$ appear as entries in $B(x_1)$.
Since multi-dimensional slices are generally described as a product of one-dimensional slices, the multi-dimensional version of the binary tensor $S(x_1,x_2)$ completely factorizes into
\begin{equation}
    S(x_1,\dots,x_{2N}) = S(x_1,x_{N+1}) \dots S(x_N,x_{2N}),
\end{equation}
so that multi-dimensional slicing operation can be described as
\begin{equation}
    B(x_1,\dots,x_N) = \sum_{x_{N+1},\dots,x_{2N}} S(x_1,x_{N+1}) \dots S(x_{N},x_{2N}) A(x_{N+1},\dots,x_{2N}).
\end{equation}
Since binary tensors based on linear functions can be efficiently converted to tensor trains and equation~\eqref{eq:slice_eq} is such a linear function, it is straightforward to construct $S(x_1,\dots,x_{2N})$ as a tensor train.
Examples can be seen in the top part of Figure \ref{fig:slicing}.

Assignment operations based on slices use the indices of a slice to define the positions of the tensor that should be assigned to some new values.
By introducing the binary tensor $T$
\begin{equation}
    T(x) = \left\liv x \mod{\slstep} = \slstart \mod{\slstep} \right\riv \left\liv \slstart \leq x < \slstop \right\riv,
    \label{eq:slice_inverse}
\end{equation}
that describes a vector where the values at the slice indices are one while all other values are zero, an assignment operation takes the following form:
\begin{align}
    B(x_1,\dots,x_N) =& \left( 1-T(x_1)\dots T(x_N) \right) B(x_1,\dots,x_N) \nonumber\\
                      &+ \sum_{x_{N+1},\dots,x_{2N}} S(x_1,x_{N+1}) \dots S(x_N,x_{2N}) A(x_{N+1},\dots,x_{2N}).
\end{align}
Here the values of $B(x_1,\dots,x_N)$ at the indices defined by multiple slices are set to the values of $A(x_{N+1},\dots,x_{2N})$.
Since equation~\eqref{eq:slice_inverse} is a combination of a modulo function and a range-based function (both defined in Section \ref{sec:general_algorithm}), it can be used for the efficient construction of tensor trains.
Examples of decompositions that represent equation~\eqref{eq:slice_inverse} are shown in the bottom part of Figure~\ref{fig:slicing}.

\paragraph{Discrete Wavelet transforms}
Discrete wavelet transforms are unitary transformations that are often used in signal processing and data compression~\cite{Guo_2022}.
Decomposition schemes have been known for a long time~\cite{Fijany_1999} and are especially interesting for the use in quantum computing~\cite{Bagherimehrab_2024}.
A discrete wavelet transform $W(x_1,x_2)$ with even $\dim(x_1)=\dim(x_2)$ is based on $M$ filter coefficients $c(m)$, which have the properties
\begin{equation}
    \sum^M_{m=1} c_m = \sqrt{\frac{1}{2}}, \ \ \ \ \ \ \ \
    \sum^M_{m=1} c_m^2 = 1.
\end{equation}
For example, when $M=4$, the matrix representation of $W(x_1,x_2)$ is
\begin{equation}
    W(x_1,x_2) = \left(\begin{array}{ccccccccccc}
    c_1&c_2&c_3&c_4&0&0&\dots&0&0&0&0 \\
    0&0&c_1&c_2&c_3&c_4&\dots&0&0&0&0 \\
    0&0&0&0&c_1&c_2&\dots&0&0&0&0 \\
    \vdots&\vdots&\vdots&\vdots&\vdots&\vdots&\ddots&\vdots&\vdots&\vdots&\vdots\\
    0&0&0&0&0&0&\dots&c_1&c_2&c_3&c_4 \\
    c_1&c_2&0&0&0&0&\dots&0&0&c_3&c_4 \\
    c_4&-c_3&c_2&-c_1&0&0&\dots&0&0&0&0 \\
    0&0&c_4&-c_3&c_2&-c_1&\dots&0&0&0&0 \\
    0&0&0&0&c_4&c_3&\dots&0&0&0&0 \\
    \vdots&\vdots&\vdots&\vdots&\vdots&\vdots&\ddots&\vdots&\vdots&\vdots&\vdots\\
    0&0&0&0&0&0&\dots&c_4&-c_3&c_2&-c_1 \\
    c_2&-c_1&0&0&0&0&\dots&0&0&c_4&-c_3 \\
    \end{array} \right).
\end{equation}
A key structural property of this operator is that the upper and lower halves share the same pattern of coefficients, differing only by sign changes.
To exploit this, we factor the index $x_1$ by 2, effectively splitting the matrix into an upper and a lower block.
Using the rank product, this yields
\begin{equation}
    W(x_1,x_2) = W(x_{1,1},x_{1,2},x_2) = P_1(x_{1,1}) \boxtimes U(x_{1,2},x_2) + P_2(x_{1,1}) \boxtimes L(x_{1,2},x_2),
    \label{eq:wavelet}
\end{equation}
with $\text{dim}(x_{1,1})=2$ and the selector tensors $P_1$ for $\dot{x}=(0,)$ and $P_2$ for $\dot{x}=(1,)$.
$U$ and $L$ can be expressed as
\begin{align}
    U(x_{1,2},x_2) &= \sum_z V(x_{1,2},x_2,x_3)\ c(x_3) \\
    L(x_{1,2},x_2) &= \sum_z V(x_{1,2},x_2,x_3)\ (-1)^{x_3} c(\text{dim}(x_3)-x_3-1)
\end{align}
where the auxiliary three-dimensional tensor $V(x_{1,2},x_2,x_3)$ is binary and defined by
\begin{figure}[b!]
    \begin{minipage}[c]{.5\textwidth}% or {.6\linewidth}, it's the same here
        \includegraphics[width=\linewidth]{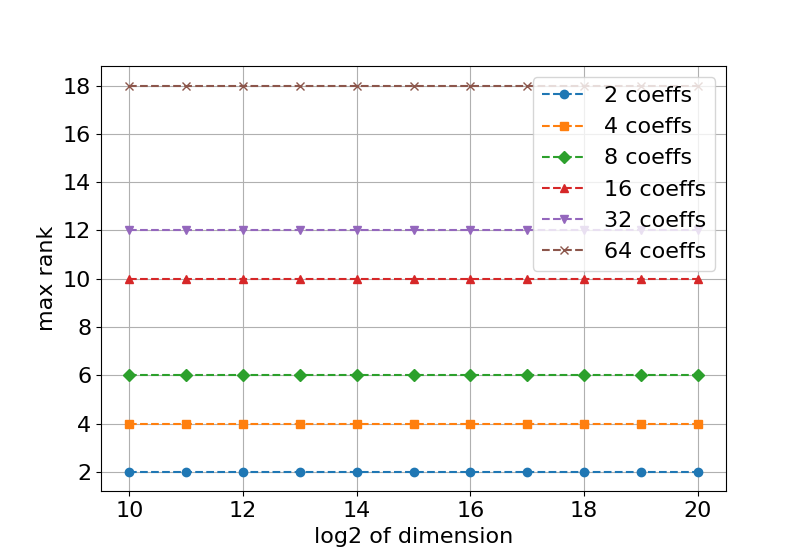}
    \end{minipage}
    \hfill
    \begin{minipage}[c]{.5\textwidth}
        \begin{tabular}{c|cccccc}
            & \multicolumn{6}{c}{number of coefficients} \\
            size & 10 & 25 & 50 & 100 & 200 & 500 \\ \hline
            $2\times 2^{15}$ &  6 & 12 & 14 & 28 & 28 & 60 \\
            $2\times3^{15}$  & 10 & 14 & 16 & 26 & 48 & 50 \\
            $2\times4^{15}$  &  8 & 10 & 16 & 20 & 30 & 44 \\
            $2\times5^{15}$  &  8 & 14 & 24 & 26 & 26 & 44 \\
        \end{tabular}
    \end{minipage}
    \caption{Left, the maximum rank of tensor train wavelet transforms are plotted against the size of the matrix $(2^i,2^i)$ for different numbers of filter coefficients. Right, the maximum ranks are tabulated for selected matrix sizes and coefficients. The results are exact without any truncation and therefore independent on the coefficients used. As can be seen, the rank is independent on the matrix size and does not increase exponentially with the number of coefficient. Depending on the filter coefficients, it might be possible to reduce the rank considerably with a subsequent truncation step.}
    \label{tab:wavelet}
\end{figure}      
\begin{equation}
    V(x_{1,2},x_2,x_3) = \left\liv 2x_{1,2}-x_2+x_3 = 0 \right\riv + \left\liv 2x_{1,2}-x_2+x_3 = \text{dim}(x_2) \right\riv.
\end{equation}
This tensor ensures that the filter coefficients are placed at the correct matrix entries. In the upper block the coefficients appear in their original order, while in the lower block they are reversed and modified by alternating signs.

Figure~\ref{tab:wavelet} displays the maximal tensor train ranks for several exactly constructed wavelet transforms, for varying numbers of filter coefficients $c_m$ and different matrix sizes.
The results show that the rank is independent of the transform size and grows only slowly with the number of filter coefficients.
Consequently, wavelet transforms, similarly to the quantum Fourier transform, can be efficiently constructed and used as fundamental building blocks in tensor train based algorithms.

%% file: sections/conclusion.tex
\section{Conclusion} \label{sec:conclusion}
In this work, we presented an algorithm based on the rank product introduced in~\cite{Kazeev_2012} for the exact factorization of binary tensors defined by some Boolean function into quantics tensor trains.
The algorithm can be applied reveal the rank structure without suffering from unfavorable scaling or convergence issues that affect SVD- or cross-interpolation–based methods, making it a useful complementary construction technique.
It is, however, not a black-box method: efficient use requires some structural knowledge of the defining Boolean function.
In particular, one needs fast procedures to test, over finite index ranges, whether the function ever evaluates to true and whether shifted versions of the same function are linearly dependent. 
We considered linear, modulo, and range-based functions as classes with good heuristics for these checks.
More general cases, such as higher-order Diophantine equations, were not addressed here but could be explored in the future to generate binary tensors with more complex patterns.

The investigated functions have been used for some selected applications, including the reordering of factorized indices, slicing and assignment operations of quantics tensor trains, discrete wavelet transforms and discrete multi-dimensional convolutions.
In particular, assignment operations are an interesting use case, since it is a missing part in libraries that deal with N-dimensional quantics tensor trains, and might enable the advancement of these libraries towards an API that is similar to other multidimensional array libraries like NumPy or PyTorch.